\documentclass[pra,onecolumn, showpacs, showkeys, secnumarabic, aps, amsmath, amssymb, nofootinbib, superscriptaddress, longbibliography, floatfix, table-of-contents, dblfloatfix]{revtex4-2}

\usepackage[utf8]{inputenc}
\usepackage[pdftex]{graphicx}
\usepackage{mathrsfs}
\usepackage[colorlinks, breaklinks, urlcolor={blue}, linkcolor={blue}, citecolor={blue}]{hyperref}
\usepackage{array}
\usepackage{amsmath}
\usepackage{type1cm}
\usepackage{lettrine}
\usepackage[english]{babel}
\usepackage{lmodern}
\usepackage{microtype}
\usepackage{booktabs}
\usepackage{caption}
\usepackage{braket}
\usepackage{xcolor}
\usepackage{orcidlink}

\begin{document}
\title{Teleporting single qutrit using symmetric-anti symmetric two-qutrit basis states as quantum channels}

\author{Anushree Pandey \orcidlink{0000-0002-0655-1773} }
\affiliation{Department of Mathematics, Techno Main Salt Lake (Engg. Colg.), \\Techno India Group, EM 4/1, Sector V, Salt Lake, Kolkata  700091, India}

\author{Abhijit Mandal\orcidlink{0000-0001-7101-9495}}
\affiliation{Department of Mathematics, Techno Main Salt Lake (Engg. Colg.), \\Techno India Group, EM 4/1, Sector V, Salt Lake, Kolkata  700091, India}

\author{Sovik Roy \orcidlink{0000-0003-4334-341X} }
\affiliation{Department of Mathematics, Techno Main Salt Lake (Engg. Colg.), \\Techno India Group, EM 4/1, Sector V, Salt Lake, Kolkata  700091, India}
\affiliation{Centre of Advanced Studies and Innovation Lab (CASILAB), 18/27, Tarapur, Silchar 788003, INDIA}

\begin{abstract}
\noindent This work presents a deterministic protocol for the teleportation of a single qutrit using maximally entangled two-qutrit states constructed from symmetric and anti-symmetric bases. Leveraging the higher information capacity of qutrits compared to qubits, we employ nine distinct two-qutrit entangled channels derived from the Leslie–Devin–Lynn (LDL) construction. For each channel, explicit unitary operations at the receiver Bob's end are derived, ensuring perfect recovery of the unknown qutrit state, initially in possesion of the sender Alice, after the sender performs joint measurements on her qutrits and communicates the outcomes through a classical channel. The proposed framework confirms teleportation with unit fidelity, thereby extending conventional teleportation schemes beyond qubits into higher-dimensional systems. This not only enriches the available toolkit for quantum communication but also highlights the utility of structured entangled states in advancing quantum information processing. The results open new avenues for secure and efficient quantum networks, higher-dimensional cryptographic schemes, and the design of novel quantum algorithms, while laying the groundwork for experimental realizations of deterministic qutrit teleportation.
\end{abstract}

\keywords{Two-qutrit states, single qutrit, entanglement, teleportation, unitary operator}
\pacs{03.67.-a, 03.67. Hk, 03.67. Bg}

\maketitle
\section{Introduction:}\label{sec:introduction}
\noindent Quantum teleportation, a cornerstone of quantum information science, has been extensively studied for two-qubits \cite{rigolin2005quantum,zhang2006experimental,pirandola2015advances} as well as for multi-qubits \cite{dong2008controlled,zhang2005many,song2018quantum,zhou2019bidirectional}. It enables the transmission of quantum states over arbitrary distances without physically transferring the quantum system itself. However, extending this principle to higher-dimensional systems involving qutrits offers potential advantages in terms of information capacity and security \cite{luo2019quantum,huang2020quantum,roy2025unitary}. Prior works on qutrit teleportation have explored various approaches, including those utilizing complex entangled states or multiple communication channels. The basic teleportation protocol \cite{bennett1993teleporting} revolves around four fundamental steps. (a) The sender Alice ($A$) and the recipient Bob ($B$) must share a special quantum link called the entangled state (or resource), which can be visualized in two different ways: either the parties interacted with one another sometimes before, or some other party prepared the entangled states for Alice and Bob, thereby sending one qubit to Alice and another to Bob. (b) Also, sender Alice holds an unknown quantum state (or qubit) which she expects to send to the receiver Bob, in which case she clubs her unknown qubit with the qubit of the entangled resource state, which she shares with Bob. (c) Alice then measures joint measurement on both the qubits she is holding. (d) Alice transmits the result of her joint measurement to Bob via a classical channel. (e) Bob, after knowing about these outcomes obtained by Alice, implements a suitable unitary operator to retrieve (or reconstruct) the unknown state Alice was initially holding, but in the due process, the original state Alice was possessing gets destroyed. The steps (a) - (e) are basic fundamental steps which are followed (theoretically) in an teleportation protocol. It is also to be noted that, the teleportation protocol is not entirely quantum in the sense that some sort of classical communication is involved here, which can be a simple telephone call or an email. This process effectively teleports the original quantum state from Alice to Bob, even though there is no traveling of physical particles, and the teleportation protocol does not violate causality either.\\\\

\noindent However, the study of quantum teleportation in higher-dimensional systems, such as qutrits (which we sometimes refer to as quantum systems with three states, viz. Horizontal polarization $H$, Vertical polarization ($V$) and Diagonal polarization ($D$) of the particles), offers several significant advantages over traditional qubit-based teleportation. Qutrits can encode more information per unit time than qubits. This higher information density translates to increased bandwidth in quantum communication channels, enabling faster and more efficient transmission of quantum data. Moreover, exploring teleportation in higher-dimensional systems can lead to the development of novel quantum algorithms and protocols that may outperform their qubit-based counterparts in various applications, such as quantum computing and quantum cryptography. Studying qutrit teleportation provides deeper insights into the fundamental principles of quantum mechanics and the nature of quantum information. It allows one to investigate the unique properties and limitations of higher-dimensional quantum systems and helps to explore the boundaries of quantum information processing.\\\\

\noindent Keeping these motivations in mind, in this article, we have considered a basis of symmetric and anti-symmetric two-qutrit states as quantum teleportation channels and have shown that a single qutrit can be teleported from Alice ($A$) to Bob ($B$) successfully. The paper is arranged as follows. We discuss symmetric and anti-symmetric basis states in section $II$. In section $III$, we show in detail how successful teleportation can be performed using the basis states defined in section $II$. This is followed by a conclusion and future directions in section $IV$.
\section{Symmetric and Anti-symmetric basis states:}
\noindent In quantum mechanics, a qutrit is a three-level quantum system, a generalization of the qubit (two-level quantum system). When dealing with multiple identical qutrits, the concept of symmetric and antisymmetric basis states becomes crucial. Symmetric States remain unchanged when two qutrits are exchanged. Mathematically, if $\vert \Psi\rangle$ is a symmetric state of two qutrits, then $\vert \Psi_{AB} \rangle= \vert \Psi_{BA}\rangle$,  where $\vert \Psi_{AB}\rangle$ represents the state of the system with qutrit $A$ in the first position and qutrit $B$ in the second position. Anti-symmetric states change sign when two-qutrits are exchanged. Mathematically, if $\vert \Phi_{AB}\rangle$ is an anti-symmetric state of two qutrits, then $\vert \Phi_{AB}\rangle=\vert \Psi_{BA}\rangle$. Thus given two single-qutrit states $\vert A\rangle$ and $\vert B\rangle$, the symmetric and antisymmetric two-qutrit states can be constructed as
\begin{eqnarray}
\vert \Psi_{AB}\rangle &=& \vert i_A\rangle \vert j_B\rangle + \vert j_B\rangle\vert i_A\rangle,\nonumber\\
\vert \Phi_{AB}\rangle &=& \vert i_A\rangle\vert j_B\rangle - \vert j_B\rangle\vert i_A\rangle.
\end{eqnarray}
where the $\vert \Psi_{AB}\rangle$ and $\vert \Phi_{AB}\rangle$ are symmetric and anti-symmetric representation. Also the plus (+) sign applies to bosons and the minus (-) sign to fermions. The concept of symmetric and anti-symmetric states extends to systems with more than two qutrits. The symmetry or anti-symmetry of a state has significant implications for the physical properties of the system and its interactions. These concepts are crucial in understanding the behavior of identical particles in quantum systems, such as in quantum information processing and condensed matter physics. The actual construction and properties of symmetric and anti-symmetric qutrit states can be more complex depending on the specific physical system and the interactions between the qutrits. With implicit symmetrization or anti-symmetrization, Leslie-Devin-Lynn (LDL) \textit{et al} \cite{leslie2019maximal} \footnote{This artilce is actually a thesis article.} had constructed the following qutrit states, which we shall consider as quantum channels for teleporting a single qutrit from sender Alice to the receiver Bob. The nine two-qutrit LDL basis states are defined as follows:
\begin{center}
\begin{subequations}
\begin{align}
 \vert \Phi^{0}_{0AB}\rangle &= \frac{1}{\sqrt{3}}\Big[ \vert H_AH_B \rangle + \vert V_AV_B\rangle + \vert D_AD_B\rangle\Big], \label{eq:sys:a} \\
 \vert \Phi^{1}_{0AB}\rangle &= \frac{1}{\sqrt{3}}\Big [\vert H_AH_B\rangle +e^{i\frac{2\pi}{3}}\vert V_AV_B\rangle + e^{i\frac{4\pi}{3}} \vert D_AD_B\rangle\Big], \label{eq:sys:b} \\
 \vert \Phi^{2}_{0AB}\rangle &= \frac{1}{\sqrt{3}}\Big [\vert H_AH_B\rangle + e^{i\frac{4\pi}{3}} \vert V_AV_B\rangle + e^{i\frac{2\pi}{3}} \vert D_AD_B\rangle\Big],\label{eq:sys:c} \\
\vert \Phi^{0}_{1AB}\rangle &= \frac{1}{\sqrt{3}}\Big[ \vert H_AV_B\rangle + \vert V_AD_B\rangle + \vert H_AD_B\rangle\Big],\label{eq:sys:d} \\
\vert \Phi^{1}_{1AB}\rangle &= \frac{1}{\sqrt{3}}\Big [\vert H_AV_B\rangle + e^{i\frac{2\pi}{3}}\vert V_AD_B\rangle + e^{i\frac{4\pi}{3}}\vert H_AD_B\rangle_{AB}\Big], \label{eq:sys:e} \\
\vert \Phi^{2}_{1AB}\rangle &= \frac{1}{\sqrt{3}}\Big [\vert H_AV_B\rangle + e^{i\frac{4\pi}{3}}\vert V_AD_B\rangle + e^{i\frac{2\pi}{3}} \vert H_AD_B\rangle_{AB}\Big], \label{eq:sys:f}\\
\vert \Phi^{0}_{2AB}\rangle &= \frac{1}{\sqrt{3}}\Big [\vert H_AD_B\rangle + \vert H_AV_B\rangle + \vert V_AD_B\rangle],\label{eq:sys:g}\\
\vert \Phi^{1}_{2AB}\rangle &= \frac{1}{\sqrt{3}}\Big [\vert H_AD_B\rangle + e^{i\frac{2\pi}{3}}\vert H_AV_B\rangle + e^{i\frac{4\pi}{3}} \vert V_AD_B\rangle\Big],\label{eq:sys:h}\\
\vert \Phi^{2}_{2AB}\rangle &= \frac{1}{\sqrt{3}}\Big [\vert H_AD_B\rangle + e^{i\frac{4\pi}{3}}\vert H_AV_B\rangle + e^{i\frac{2\pi}{3}} \vert V_AD_B\rangle\Big],\label{eq:sys:i}
\end{align}
\end{subequations}
\end{center}
where, $\Phi^{i}_{jAB}$ ($i = 0,\: 1,\:2$ and $j = 0,\: 1,\:2$) denotes that the two qutrit states are being shared between Alice ($A$) and Bob ($B$). On the right hand side $X_AY_B$ represents that the respective qutrit $X$ is in Alice's possession and $Y$ is in Bob's possession. Also $X,\: Y = H,\: V,\: D$. The channels described by equations (\ref{eq:sys:a})-(\ref{eq:sys:i}) are maximally entangled qutrit states with both symmetric and anti-symmetric properties. These states are chosen because they form a complete basis, which is essential for deterministic quantum teleportation. Leveraging the higher information capacity of qutrits over qubits, these channels allow for more efficient quantum communication. They enable a sender (Alice) to perform joint measurements and a receiver (Bob) to apply a specific unitary correction to perfectly reconstruct the original qutrit. This protocol exploits the inherent symmetry of the two-qutrit system, ensuring high-fidelity teleportation and making these channels optimal for advanced quantum protocols.
\section{Teleportation of single qutrit via two qutrit basis channels}
\noindent Alice ($A$) holds a single qutrit which she wants to send to Bob ($B$). The single qutrit held by Alice is
\begin{eqnarray}
\label{singlequtrit}
\vert \Psi_A\rangle = \alpha\vert H_A\rangle + \beta\vert V_A\rangle + \gamma\vert D_A\rangle.
\end{eqnarray}
Alice will consider the states (\ref{eq:sys:a})-(\ref{eq:sys:i}) as quantum channels for performing standard teleportation. Also, we can re-write the Eqs.(\ref{eq:sys:a})-(\ref{eq:sys:i}), as
\begin{eqnarray}
\label{rewritebasis}
\vert H_AH_B\rangle &=& \frac{1}{\sqrt{3}}\Big(\vert \Phi^0_{0AB}\rangle + \vert \Phi^1_{0AB}\rangle + \vert \Phi^2_{0AB}\rangle\Big),\nonumber\\
\vert H_AV_B\rangle &=& \frac{1}{\sqrt{3}}\Big(\vert \Phi^0_{1AB}\rangle + \vert \Phi^1_{1AB}\rangle + \vert \Phi^2_{1AB}\rangle\Big),\nonumber\\
\vert H_AD_B\rangle &=& \frac{1}{\sqrt{3}}\Big(\vert \Phi^0_{2AB}\rangle + \vert \Phi^1_{2AB}\rangle + \vert \Phi^2_{2AB}\rangle\Big),\nonumber\\
\vert V_AH_B\rangle &=& \frac{1}{\sqrt{3}}\Big(\vert \Phi^0_{2AB}\rangle + e^{4\frac{\pi\:i}{3}}\vert \Phi^1_{2AB}\rangle + e^{2\frac{\pi\:i}{3}}\vert \Phi^2_{2AB}\rangle\Big),\nonumber\\
\vert V_AV_B\rangle &=& \frac{1}{\sqrt{3}}\Big(\vert \Phi^0_{0AB}\rangle + e^{4\frac{\pi\:i}{3}}\vert \Phi^1_{0AB}\rangle + e^{2\frac{\pi\:i}{3}}\vert \Phi^2_{0AB}\rangle\Big),\nonumber\\
\vert V_AD_B\rangle &=& \frac{1}{\sqrt{3}}\Big(\vert \Phi^0_{1AB}\rangle+ e^{4\frac{\pi\:i}{3}}\vert \Phi^1_{1AB}\rangle + e^{2\frac{\pi\:i}{3}}\vert \Phi^2_{1AB}\rangle\Big),\nonumber\\
\vert D_AH_B\rangle &=& \frac{1}{\sqrt{3}}\Big(\vert \Phi^0_{1AB}\rangle + e^{2\frac{\pi\:i}{3}}\vert \Phi^1_{1AB}\rangle + e^{4\frac{\pi\:i}{3}}\vert \Phi^2_{1AB}\rangle\Big),\nonumber\\
\vert D_AV_B\rangle &=& \frac{1}{\sqrt{3}}\Big(\vert \Phi^0_{2AB}\rangle + e^{2\frac{\pi\:i}{3}}\vert \Phi^1_{2AB}\rangle + e^{4\frac{\pi\:i}{3}}\vert \Phi^2_{2AB}\rangle\Big),\nonumber\\
\vert D_AD_B\rangle &=& \frac{1}{\sqrt{3}}\Big(\vert \Phi^0_{0AB}\rangle + e^{2\frac{\pi\:i}{3}}\vert \Phi^1_{0AB}\rangle + e^{4\frac{\pi\:i}{3}}\vert \Phi^2_{0AB}\rangle\Big).\nonumber\\
\end{eqnarray}
\noindent We shall now discuss the teleportation process by highlighting the important phases, considering the cases group-wise, which means out of the above nine LDL states, some states, when used as a shared resource, require the same unitary transformation from Bob's end.
\subsection{The states $\vert \Phi^{0}_{0AB}\rangle,\:\vert \Phi^{1}_{0AB}\rangle,\:\vert \Phi^{2}_{0AB}\rangle$ as teleportation channels:}
\noindent When Alice uses Eq.(\ref{eq:sys:a})-(\ref{eq:sys:c}) as quantum channels for teleportation, she clubs her single qutrit state (\ref{singlequtrit}) to the channels and restructure the system in such a way that she now holds to qutrits in her possession. Using Eqs. (\ref{rewritebasis}) the following are obtained.
\begin{eqnarray}
\label{clubbedstate1}
\vert \Phi^{0}_{0AAB}\rangle^{cl} &=& \frac{1}{3}\Big(\vert \Phi^0_{0AA}\rangle (\alpha\vert H_B\rangle + \beta\vert V_B\rangle + \gamma\vert D_B\rangle) + \vert \Phi^1_{0AA}\rangle (\alpha\vert H_B\rangle + e^{4\frac{\pi\:i}{3}}\beta\vert V_B\rangle + e^{2\frac{\pi\:i}{3}}\gamma\vert D_B\rangle)\nonumber\\ && +\vert \Phi^2_{0AA}\rangle (\alpha\vert H_B\rangle + e^{2\frac{\pi\:i}{3}}\beta\vert V_B\rangle + e^{4\frac{\pi\:i}{3}}\gamma\vert D_B\rangle) + \vert \Phi^0_{1AA}\rangle (\alpha\vert V_B\rangle + \beta\vert D_B\rangle + \gamma\vert H_B\rangle) \nonumber\\ && + \vert \Phi^1_{1AA}\rangle (\alpha\vert V_B\rangle + e^{4\frac{\pi\:i}{3}}\beta\vert D_B\rangle + e^{2\frac{\pi\:i}{3}}\gamma\vert H_B\rangle)+ \vert \Phi^2_{1AA}\rangle (\alpha\vert V_B\rangle + e^{2\frac{\pi\:i}{3}}\beta\vert D_B\rangle + e^{4\frac{\pi\:i}{3}}\gamma\vert H_B\rangle) \nonumber\\ && + \vert \Phi^0_{2AA}\rangle (\alpha\vert D_B\rangle + \beta\vert H_B\rangle + \gamma\vert V_B\rangle) + \vert \Phi^1_{2AA}\rangle (\alpha\vert D_B\rangle + e^{4\frac{\pi\:i}{3}}\beta\vert H_B\rangle + e^{2\frac{\pi\:i}{3}}\gamma\vert V_B\rangle) \nonumber\\ && + \vert \Phi^2_{2AA}\rangle (\alpha\vert D_B\rangle + e^{2\frac{\pi\:i}{3}}\beta\vert H_B\rangle + e^{4\frac{\pi\:i}{3}}\gamma\vert V_B\rangle)\Big),
\end{eqnarray}
\begin{eqnarray}
\label{clubbedstate2}
\vert \Phi^{1}_{0}\rangle^{cl}_{AAB} &=& \frac{1}{3}\Big(\vert \Phi^0_{0AA}\rangle (\alpha\vert H_B\rangle + e^{2\frac{\pi\:i}{3}}\beta\vert V_B\rangle + e^{4\frac{\pi\:i}{3}}\gamma\vert D_B\rangle) + \vert \Phi^1 \rangle (\alpha\vert H_B\rangle + e^{2\pi\:i}\beta\vert V_B\rangle + e^{2\pi\:i}\gamma\vert D_B\rangle)\nonumber\\ && +\vert \Phi^2 \rangle (\alpha\vert H_B\rangle + e^{4\frac{\pi\:i}{3}}\beta\vert V_B\rangle + e^{8\frac{\pi\:i}{3}}\gamma\vert D_B\rangle) + \vert \Phi^0_{1AA}\rangle (e^{2\frac{\pi\:i}{3}}\alpha\vert V_B\rangle + e^{4\frac{\pi\:i}{3}}\beta\vert D_B\rangle + \gamma\vert H_B\rangle) \nonumber\\ && + \vert \Phi^1_{1AA}\rangle (e^{2\frac{\pi\:i}{3}}\alpha\vert V_B\rangle + e^{8\frac{\pi\:i}{3}}\beta\vert D_B\rangle + e^{2\frac{\pi\:i}{3}}\gamma\vert H_B\rangle)+ \vert \Phi^2_{1AA}\rangle (e^{2\frac{\pi\:i}{3}}\alpha\vert V_B\rangle + e^{2\pi\:i}\beta\vert D_B\rangle + e^{4\frac{\pi\:i}{3}}\gamma\vert H_B \rangle) \nonumber\\ && + \vert \Phi^0_{2AA}\rangle (e^{4\frac{\pi\:i}{3}}\alpha\vert D_B\rangle + \beta\vert H_B\rangle + e^{2\frac{\pi\:i}{3}}\gamma\vert V_B\rangle) + \vert \Phi^1_{2AA}\rangle (e^{4\frac{\pi\:i}{3}}\alpha\vert D_B\rangle + e^{4\frac{\pi\:i}{3}}\beta\vert H_B\rangle + e^{4\frac{\pi\:i}{3}}\gamma\vert V_B\rangle) \nonumber\\ && + \vert \Phi^2_{2AA}\rangle (e^{4\frac{\pi\:i}{3}}\alpha\vert D_B\rangle + e^{2\frac{\pi\:i}{3}}\beta\vert H_B\rangle + e^{2\pi\:i}\gamma\vert V_B\rangle)\Big),
\end{eqnarray}
and
\begin{eqnarray}
\label{clubbedstate3}
\vert \Phi^{2}_{0AAB}\rangle^{cl} &=& \frac{1}{3}\Big(\vert \Phi^0_{0AA}\rangle (\alpha\vert H_B\rangle + e^{4\frac{\pi\:i}{3}}\beta\vert V_B\rangle + e^{2\frac{\pi\:i}{3}}\gamma\vert D_B\rangle) + \vert \Phi^1_{0AA}\rangle (\alpha\vert H_B\rangle + e^{8\frac{\pi\:i}{3}}\beta\vert V_B\rangle + e^{4\frac{\pi\:i}{3}}\gamma\vert D_B\rangle)\nonumber\\ && +\vert \Phi^2_{0AA}\rangle (\alpha\vert H_B\rangle + e^{2\pi\:i}\beta\vert V_B\rangle + e^{2\pi\:i}\gamma\vert D_B\rangle) + \vert \Phi^0_{1AA}\rangle (e^{4\frac{\pi\:i}{3}}\alpha\vert V_B\rangle + e^{2\frac{\pi\:i}{3}}\beta\vert D_B\rangle + \gamma\vert H_B\rangle) \nonumber\\ && + \vert \Phi^1_{1AA}\rangle (e^{4\frac{\pi\:i}{3}}\alpha\vert V_B\rangle + e^{2\pi\:i}\beta\vert D_B\rangle + e^{2\frac{\pi\:i}{3}}\gamma\vert H_B\rangle)+ \vert \Phi^2_{1AA}\rangle (e^{4\frac{\pi\:i}{3}}\alpha\vert V_B\rangle + e^{4\frac{\pi\:i}{3}}\beta\vert D_B\rangle + e^{4\frac{\pi\:i}{3}}\gamma\vert H_B\rangle) \nonumber\\ && + \vert \Phi^0_{2AA}\rangle (e^{2\frac{\pi\:i}{3}}\alpha\vert D_B\rangle + \beta\vert H_B\rangle + e^{4\frac{\pi\:i}{3}}\gamma\vert V_B\rangle) + \vert \Phi^1_{2AA}\rangle (e^{2\frac{\pi\:i}{3}}\alpha\vert D_B\rangle + e^{4\frac{\pi\:i}{3}}\beta\vert H_B\rangle + e^{2\pi\:i}\gamma\vert V_B\rangle) \nonumber\\ && + \vert \Phi^2_{2AA}\rangle (e^{2\frac{\pi\:i}{3}}\alpha\vert D_B\rangle + e^{2\frac{\pi\:i}{3}}\beta\vert H_B\rangle + e^{8\frac{\pi\:i}{3}}\gamma\vert V_B\rangle)\Big).
\end{eqnarray}
Now for each of the above cases shown as Eqs.(\ref{clubbedstate1})-(\ref{clubbedstate3}), two qutrits are possessed by Alice and a single qutrit is now being retained by Bob. Alice makes joint measurements on both her qutrits and communicates her results to Bob. Consequently, Bob applies a set of unitary operators to retrieve the original single qutrit given in Eq.(\ref{singlequtrit}). Respectively, with respect to the Eqs.(\ref{clubbedstate1})-(\ref{clubbedstate3}), the unitary operators that Bob uses are $\Big\lbrace( U_{0}^{\prime})_{i}\Big\rbrace$, $\Big\lbrace( U_{0}^{\prime\prime})_{i}\Big\rbrace$, and $\Big\lbrace( U_{0}^{\prime\prime\prime})_{i}\Big\rbrace$, where $i = 0,1,\cdots, 8$. We get the following.
\subsubsection*{When $\vert \Phi^{0}_{0AB}\rangle$ is the qutrit channel:}
\noindent In the table~$I$ we describe in the two columns what Alice obtains as outcomes after joint measurements (left column) and what Bob needs to do (right column) to retrieve the unknown qutrit defined in Eq.(\ref{singlequtrit}).

\begin{table}[h]
\caption{When the quantum channel is the state $\vert \Phi^{0}_{0AB}\rangle$}\label{tab1}%
\begin{tabular}{@{}ll@{}}
\toprule
\textbf{Measurement outcomes by Alice} & \textit{Unitary Operator that $B$ applies}\\
\midrule
$\vert \Phi^0_{0AA}\rangle$ & $(U_{0}^{\prime})_{0}\:(\alpha\vert H_B\rangle + \beta\vert V_B\rangle + \gamma\vert D_B\rangle)$\\
$\vert \Phi^1_{0AA}\rangle$ & $(U_{0}^{\prime})_{1}\:(\alpha\vert H_B\rangle + e^{4\frac{\pi\:i}{3}}\beta\vert V_B\rangle + e^{2\frac{\pi\:i}{3}}\gamma\vert D_B\rangle)$\\
$\vert \Phi^2_{0AA}\rangle$ & $(U_{0}^{\prime})_{2}\:(\alpha\vert H_B\rangle + e^{2\frac{\pi\:i}{3}}\beta\vert V_B\rangle + e^{4\frac{\pi\:i}{3}}\gamma\vert D_B\rangle)$\\
$\vert \Phi^0_{1AA}\rangle$ & $(U_{0}^{\prime})_{3}\:(\alpha\vert V_B\rangle + \beta\vert D_B\rangle + \gamma\vert H_B\rangle)$\\
$\vert \Phi^1_{1AA}\rangle$ & $(U_{0}^{\prime})_{4}\:(\alpha\vert V_B\rangle + e^{4\frac{\pi\:i}{3}}\beta\vert D_B\rangle + e^{2\frac{\pi\:i}{3}}\gamma\vert H_B\rangle)$\\
$\vert \Phi^2_{1AA}\rangle$ & $(U_{0}^{\prime})_{5}\:(\alpha\vert V_B\rangle + e^{2\frac{\pi\:i}{3}}\beta\vert D_B\rangle + e^{4\frac{\pi\:i}{3}}\gamma\vert H_B\rangle)$\\
$\vert \Phi^0_{2AA}\rangle$ & $(U_{0}^{\prime})_{6}\:(\alpha\vert D_B\rangle + \beta\vert H_B\rangle + \gamma\vert V_B\rangle)$\\
$\vert \Phi^1_{2AA}\rangle$ & $(U_{0}^{\prime})_{7}\:(\alpha\vert D_B\rangle + e^{4\frac{\pi\:i}{3}}\beta\vert H_B\rangle + e^{2\frac{\pi\:i}{3}}\gamma\vert V_B\rangle)$\\
$\vert \Phi^1_{2AA}\rangle$ & $(U_{0}^{\prime})_{8}\:(\alpha\vert D_B\rangle + e^{2\frac{\pi\:i}{3}}\beta\vert H_B\rangle + e^{4\frac{\pi\:i}{3}}\gamma\vert V_B\rangle)$\\
\botrule
\end{tabular}
\end{table}

\noindent In table $I$, $(U_{0}^{\prime})_{0}$ is the identity operator, which means that if Alice gets $\vert \Phi^0_{0AA}\rangle$, then Bob is automatically left with the original single qutrit. He does nothing, or we can say Bob applies identity operator $\mathcal{I}_B$ which is denoted here by  $(U_{0}^{\prime})_{0}$. For other outcomes, however, Bob applies unitary operators on the measurement outcomes to retreive the original single qutrit (as shown in the right hand column of the table).  Hence we get
\begin{eqnarray}
\label{unitary1}
(U_{0}^{\prime})_{0} &=& \mathcal{I}_B,\nonumber\\
(U_{0}^{\prime})_{1} &=& \vert H_B\rangle\langle H_B\vert  + e^{2\frac{\pi\: i}{3}}\vert V_B\rangle\langle V_B\vert +  e^{4\frac{\pi\: i}{3}}\vert D_B\rangle\langle D_B\vert,\nonumber\\
(U_{0}^{\prime})_{2} &=& \vert H_B \rangle\langle H_B\vert  + e^{4\frac{\pi\: i}{3}}\vert V_B\rangle \langle V_B\vert +  e^{2\frac{\pi\: i}{3}}\vert D_B\rangle \langle D_B\vert,\nonumber\\
(U_{0}^{\prime})_{3} &=& \vert H_B\rangle\langle V_B\vert  + \vert V_B\rangle\langle D_B\vert +  \vert D_B\rangle\langle H_B\vert,\nonumber\\
(U_{0}^{\prime})_{4} &=& \vert H_B\rangle\langle V_B\vert  + e^{2\frac{\pi\: i}{3}}\vert V_B\rangle\langle D_B\vert +  e^{4\frac{\pi\: i}{3}}\vert D_B\rangle\langle H_B\vert,\nonumber\\
(U_{0}^{\prime})_{5} &=& \vert H_B\rangle\langle V_B\vert  + e^{4\frac{\pi\: i}{3}}\vert V_B\rangle\langle D_B\vert +  e^{2\frac{\pi\: i}{3}}\vert D_B\rangle\langle H_B\vert,\nonumber\\
(U_{0}^{\prime})_{6} &=& \vert H_B\rangle\langle D_B\vert  + \vert V_B\rangle\langle H_B\vert +  \vert D_B\rangle\langle V_B\vert,\nonumber\\
(U_{0}^{\prime})_{7} &=& \vert H_B\rangle\langle D_B\vert  + e^{2\frac{\pi\: i}{3}}\vert V_B\rangle\langle H_B\vert +  e^{4\frac{\pi\: i}{3}}\vert D_B\rangle\langle V_B\vert,\nonumber\\
(U_{0}^{\prime})_{8} &=& \vert H_B\rangle\langle D_B\vert  + e^{4\frac{\pi\: i}{3}}\vert V_B\rangle\langle H_B\vert +  e^{2\frac{\pi\: i}{3}}\vert D_B\rangle\langle V_B\vert.\nonumber\\
\end{eqnarray}
Now for other entangled two-qutrit states as resources, we shall just tabulate the outcomes of Alice and the unitary operators applied by Bob to reconstruct the original unknown state defined in Eq.(\ref{singlequtrit}).
\subsubsection*{When $\vert \Phi^{1}_{0AB}\rangle$ is the qutrit channel:}

\begin{table}[h]
\caption{When the quantum channel is the state $\vert \Phi^{1}_{0AB}\rangle$}\label{tab2}%
\begin{tabular}{@{}ll@{}}
\toprule
\textbf{Measurement outcomes by Alice} & \textit{Unitary Operator that $B$ applies}\\
\midrule
$\vert \Phi^0_{0AA}\rangle$ & $(U_{0}^{\prime\prime})_{0}\:(\alpha\vert H_B\rangle + e^{2\frac{\pi\:i}{3}}\beta\vert V_B\rangle + e^{4\frac{\pi\:i}{3}}\gamma\vert D_B\rangle)$\\
$\vert \Phi^1_{0AA}\rangle$ & $(U_{0}^{\prime\prime})_{1}\:(\alpha\vert H_B\rangle + e^{2\pi\:i}\beta\vert V_B\rangle + e^{2\pi\:i}\gamma\vert D_B\rangle)$\\
$\vert \Phi^2_{0AA}\rangle$ & $(U_{0}^{\prime\prime})_{2}\:(\alpha\vert H_B\rangle + e^{4\frac{\pi\:i}{3}}\beta\vert V_B\rangle + e^{8\frac{\pi\:i}{3}}\gamma\vert D_B\rangle)$\\
$\vert \Phi^0_{1AA}\rangle$ & $(U_{0}^{\prime\prime})_{3}\:(e^{2\frac{\pi\:i}{3}}\alpha\vert V_B\rangle + e^{4\frac{\pi\:i}{3}}\beta\vert D_B\rangle + \gamma\vert H_B\rangle)$\\
$\vert \Phi^1_{1AA}\rangle$ & $(U_{0}^{\prime\prime})_{4}\:(e^{2\frac{\pi\:i}{3}}\alpha\vert V_B\rangle + e^{8\frac{\pi\:i}{3}}\beta\vert D_B\rangle + e^{2\frac{\pi\:i}{3}}\gamma\vert H_B\rangle)$\\
$\vert \Phi^2_{1AA}\rangle$ & $(U_{0}^{\prime\prime})_{5}\:(e^{2\frac{\pi\:i}{3}}\alpha\vert V_B\rangle + e^{2\pi\:i}\beta\vert D_B\rangle + e^{4\frac{\pi\:i}{3}}\gamma\vert H_B\rangle_{B})$\\
$\vert \Phi^0_{2AA}\rangle$ & $(U_{0}^{\prime\prime})_{6}\:(e^{4\frac{\pi\:i}{3}}\alpha\vert D_B\rangle + \beta\vert H_B\rangle + e^{2\frac{\pi\:i}{3}}\gamma\vert V_B\rangle)$\\
$\vert \Phi^1_{2AA}\rangle$ & $(U_{0}^{\prime\prime})_{7}\:(e^{4\frac{\pi\:i}{3}}\alpha\vert D_B\rangle + e^{4\frac{\pi\:i}{3}}\beta\vert H_B\rangle + e^{4\frac{\pi\:i}{3}}\gamma\vert V_B\rangle)$\\
$\vert \Phi^1_{2AA}\rangle$ & $(U_{0}^{\prime\prime})_{8}\:(e^{4\frac{\pi\:i}{3}}\alpha\vert D_B\rangle + e^{2\frac{\pi\:i}{3}}\beta\vert H_B\rangle + e^{2\pi\:i}\gamma\vert V_B\rangle)$\\
\hline
\botrule
\end{tabular}
\end{table}

\noindent The right hand column indicates how unitary operator will be applied to the state to retrieve the original qutrit. In the table $II$ $(U_{0}^{\prime\prime})_{1}$ is the identity operator, which means that if Alice gets $\vert \Phi^0_{1}\rangle_{AA}$, then Bob is automatically left with the original single qutrit. He does nothing or we can say Bob applies identity operator $\mathcal{I}$ which is denoted here by  $(U_{0}^{\prime\prime})_{1}$. Hence we get
\begin{eqnarray}
\label{unitary2}
(U_{0}^{\prime\prime})_{0} &=& \vert H_B\rangle\langle H_B\vert  + e^{4\frac{\pi\: i}{3}}\vert V_B\rangle\langle V_B\vert +  e^{2\frac{\pi\: i}{3}}\vert D_B\rangle\langle D_B\vert,\nonumber\\
(U_{0}^{\prime\prime})_{1} &=& \mathcal{I}_B,\nonumber\\
(U_{0}^{\prime\prime})_{2} &=& \vert H_B\rangle\langle H_B\vert  + e^{2\frac{\pi\: i}{3}}\vert V_B\rangle\langle V_B\vert +  e^{-2\frac{\pi\: i}{3}}\vert D_B\rangle\langle D_B\vert,\nonumber\\
(U_{0}^{\prime\prime})_{3} &=& e^{4\frac{\pi\: i}{3}}\vert H_B\rangle\langle V_B\vert  + e^{2\frac{\pi\: i}{3}}\vert V_B\rangle\langle D_B\vert +  \vert D_B\rangle\langle H_B\vert ,\nonumber\\
(U_{0}^{\prime\prime})_{4} &=& e^{4\frac{\pi\: i}{3}}\vert H_B\rangle\langle V_B\vert  + e^{-2\frac{\pi\: i}{3}}\vert V_B\rangle\langle D_B\vert +  e^{4\frac{\pi\: i}{3}}\vert D_B\rangle\langle H_B\vert,\nonumber\\
(U_{0}^{\prime\prime})_{5} &=& e^{4\frac{\pi\: i}{3}}\vert H_B\rangle\langle V_B\vert  + \vert V_B\rangle\langle D_B\vert +  e^{2\frac{\pi\: i}{3}}\vert D_B\rangle\langle H_B\vert ,\nonumber\\
(U_{0}^{\prime\prime})_{6} &=& e^{2\frac{\pi\: i}{3}}\vert H_B\rangle\langle D_B\vert  + \vert V_B\rangle\langle H_B\vert +  e^{4\frac{\pi\: i}{3}}\vert D_B\rangle\langle V_B\vert,\nonumber\\
(U_{0}^{\prime\prime})_{7} &=& e^{2\frac{\pi\: i}{3}}\vert H_B\rangle\langle D_B\vert  + e^{2\frac{\pi\: i}{3}}\vert V_B\rangle\langle H_B\vert +  e^{2\frac{\pi\: i}{3}}\vert D_B\rangle\langle V_B\vert,\nonumber\\
(U_{0}^{\prime\prime})_{8} &=& e^{2\frac{\pi\: i}{3}}\vert H_B\rangle\langle D_B\vert  + e^{4\frac{\pi\: i}{3}}\vert V_B\rangle\langle H_B\vert +  \vert D_B\rangle\langle V_B\vert .\nonumber\\
\end{eqnarray}

\subsubsection*{When $\vert \Phi^{2}_{0AB}\rangle$ is the qutrit channel:}

\begin{table}[h]
\caption{When the quantum channel is the state $\vert \Phi^{2}_{0AB}\rangle$}\label{tab2}%
\begin{tabular}{@{}ll@{}}
\toprule
\textbf{Measurement outcomes by Alice} & \textit{Unitary Operator that $B$ applies}\\
\midrule
$\vert \Phi^0_{0AA}\rangle$ & $(U_{0}^{\prime\prime\prime})_{0}\:(\alpha\vert H_B\rangle + e^{4\frac{\pi\:i}{3}}\beta\vert V_B\rangle + e^{2\frac{\pi\:i}{3}}\gamma\vert D_B\rangle_{B})$\\
$\vert \Phi^1_{0AA}\rangle$ & $(U_{0}^{\prime\prime\prime})_{1}\:(\alpha\vert H_B\rangle + e^{8\frac{\pi\:i}{3}}\beta\vert V_B\rangle + e^{4\frac{\pi\:i}{3}}\gamma\vert D_B\rangle)$\\
$\vert \Phi^2_{0AA}\rangle$ & $(U_{0}^{\prime\prime\prime})_{2}\:(\alpha\vert H_B\rangle + \beta\vert V_B\rangle + \gamma\vert D_B\rangle)$\\

$\vert \Phi^0_{1AA}\rangle$ & $(U_{0}^{\prime\prime\prime})_{3}\:(e^{4\frac{\pi\:i}{3}}\alpha\vert V_B\rangle + e^{2\frac{\pi\:i}{3}}\beta\vert D_B\rangle + \gamma\vert H_B\rangle)$\\

$\vert \Phi^1_{1AA}\rangle$ & $(U_{0}^{\prime\prime\prime})_{4}\:(e^{4\frac{\pi\:i}{3}}\alpha\vert V_B\rangle + \beta\vert D_B\rangle + e^{2\frac{\pi\:i}{3}}\gamma\vert H_B\rangle)$\\

$\vert \Phi^2_{1AA}\rangle$ & $(U_{0}^{\prime\prime\prime})_{5}\:(e^{4\frac{\pi\:i}{3}}\alpha\vert V_B\rangle + e^{4\frac{\pi\:i}{3}}\beta\vert D_B\rangle + e^{4\frac{\pi\:i}{3}}\gamma\vert H_B\rangle)$\\

$\vert \Phi^0_{2AA}\rangle$ & $(U_{0}^{\prime\prime\prime})_{6}\:(e^{2\frac{\pi\:i}{3}}\alpha\vert D_B\rangle + \beta\vert H_B\rangle + e^{4\frac{\pi\:i}{3}}\gamma\vert V_B\rangle)$\\

$\vert \Phi^1_{2AA}\rangle$ & $(U_{0}^{\prime\prime\prime})_{7}\:(e^{2\frac{\pi\:i}{3}}\alpha\vert D_B\rangle + e^{4\frac{\pi\:i}{3}}\beta\vert H_B\rangle + \gamma\vert V_B\rangle)$\\

$\vert \Phi^1_{2AA}\rangle$ & $(U_{0}^{\prime\prime\prime})_{8}\:(e^{2\frac{\pi\:i}{3}}\alpha\vert D_B\rangle + e^{2\frac{\pi\:i}{3}}\beta\vert H_B\rangle + e^{8\frac{\pi\:i}{3}}\gamma\vert V_B\rangle)$\\
\hline
\botrule
\end{tabular}
\end{table}
\noindent 
In table~$III$, $(U_{0}^{\prime\prime\prime})_{2}$ is the identity operator, which means that if Alice gets $\vert \Phi^2_{0AA}\rangle$, then Bob is automatically left with the original single qutrit. He does nothing or we can say Bob applies identity operator $\mathcal{I}$ which is denoted here by  $(U_{0}^{\prime\prime\prime})_{2}$. Hence we get
\begin{eqnarray}
\label{unitary3}
(U_{0}^{\prime\prime\prime})_{0} &=& \vert H_B\rangle\langle H_B\vert  + e^{2\frac{\pi\: i}{3}}\vert V_B\rangle\langle V_B\vert +  e^{4\frac{\pi\: i}{3}}\vert D_B\rangle\langle D_B\vert,\nonumber\\
(U_{0}^{\prime\prime\prime})_{1} &=& \vert H_B\rangle\langle H_B\vert  + e^{-2\frac{\pi\: i}{3}}\vert V_B\rangle\langle V_B\vert +  e^{4\frac{\pi\: i}{3}}\vert D_B\rangle\langle D_B\vert,\nonumber\\
(U_{0}^{\prime\prime\prime})_{2} &=& \mathcal{I}_B,\nonumber\\
(U_{0}^{\prime\prime\prime})_{3} &=& e^{2\frac{\pi\: i}{3}}\vert H_B\rangle\langle V_B\vert  + e^{4\frac{\pi\: i}{3}}\vert V_B\rangle\langle D_B\vert +  \vert D_B\rangle\langle H_B\vert ,\nonumber\\
(U_{0}^{\prime\prime\prime})_{4} &=& e^{2\frac{\pi\: i}{3}}\vert H_B\rangle\langle V_B\vert  + \vert V_B\rangle\langle D_B\vert + e^{4\frac{\pi\: i}{3}} \vert D_B\rangle\langle H_B\vert,\nonumber\\
(U_{0}^{\prime\prime\prime})_{5} &=& e^{2\frac{\pi\: i}{3}}\vert H_B\rangle\langle V_B\vert  + e^{2\frac{\pi\: i}{3}}\vert V_B\rangle\langle D_B\vert + e^{2\frac{\pi\: i}{3}} \vert D_B\rangle\langle H_B\vert ,\nonumber\\
(U_{0}^{\prime\prime\prime})_{6} &=& e^{4\frac{\pi\: i}{3}}\vert H_B\rangle\langle D_B\vert  + \vert V_B\rangle\langle H_B\vert + e^{2\frac{\pi\: i}{3}} \vert D_B\rangle\langle V_B\vert,\nonumber\\
(U_{0}^{\prime\prime\prime})_{7} &=& e^{4\frac{\pi\: i}{3}}\vert H_B\rangle\langle D_B\vert  + e^{2\frac{\pi\: i}{3}}\vert V_B\rangle\langle H_B\vert + \vert D_B\rangle\langle V_B\vert ,\nonumber\\
(U_{0}^{\prime\prime\prime})_{8} &=& e^{4\frac{\pi\: i}{3}} \vert H_B\rangle\langle D_B\vert  + e^{4\frac{\pi\: i}{3}}\vert V_B\rangle\langle H_B\vert +  e^{-2\frac{\pi\: i}{3}}\vert D_B\rangle\langle V_B\vert .\nonumber\\
\end{eqnarray}
\subsection*{The states $\vert \Phi^{0}_{1AB}\rangle,\:\vert \Phi^{1}_{1AB}\rangle,\:\vert \Phi^{2}_{1AB}\rangle$ as teleportation channels:}
\noindent As before, if we now consider the states defined in Eqs.(\ref{eq:sys:d})-(\ref{eq:sys:f})as quantum channels for teleportation, then using Eqs.(\ref{rewritebasis}) the following are obtained.
\begin{eqnarray}
\label{clubbedstate4}
\vert \Phi^{0}_{1AAB}\rangle^{cl} &=& \frac{1}{3}\Big(\vert \Phi^0_{0AA}\rangle (\alpha\vert V_B\rangle + \beta\vert D_B\rangle + \gamma\vert H_B\rangle) + \vert \Phi^1_{0AA}\rangle (\alpha\vert V_B\rangle + e^{4\frac{\pi\:i}{3}}\beta\vert D_B\rangle + e^{2\frac{\pi\:i}{3}}\gamma\vert H_B\rangle)\nonumber\\ && +\vert \Phi^2_{0AA}\rangle (\alpha\vert V_B\rangle + e^{2\frac{\pi\:i}{3}}\beta\vert D_B\rangle + e^{4\frac{\pi\:i}{3}}\gamma\vert H_B\rangle) + \vert \Phi^0_{1AA}\rangle (\alpha\vert D_B\rangle + \beta\vert H_B\rangle + \gamma\vert V_B\rangle) \nonumber\\ && + \vert \Phi^1_{1AA}\rangle (\alpha\vert D_B\rangle + e^{4\frac{\pi\:i}{3}}\beta\vert H_B\rangle + e^{2\frac{\pi\:i}{3}}\gamma\vert V_B\rangle)+ \vert \Phi^2_{1AA}\rangle (\alpha\vert D_B\rangle + e^{2\frac{\pi\:i}{3}}\beta\vert H_B\rangle + e^{4\frac{\pi\:i}{3}}\gamma\vert V_B\rangle) \nonumber\\ && + \vert \Phi^0_{2AA}\rangle (\alpha\vert H_B\rangle + \beta\vert V_B\rangle + \gamma\vert D_B\rangle) + \vert \Phi^1_{2AA}\rangle (\alpha\vert H_B\rangle + e^{4\frac{\pi\:i}{3}}\beta\vert V_B\rangle + e^{2\frac{\pi\:i}{3}}\gamma\vert D_B\rangle) \nonumber\\ && + \vert \Phi^2_{2AA}\rangle (\alpha\vert H_B\rangle + e^{2\frac{\pi\:i}{3}}\beta\vert V_B\rangle + e^{4\frac{\pi\:i}{3}}\gamma\vert D_B\rangle)\Big),
\end{eqnarray}
\begin{eqnarray}
\label{clubbedstate5}
\vert \Phi^{1}_{1AAB}\rangle^{cl} &=& \frac{1}{3}\Big(\vert \Phi^0_{0AA}\rangle (\alpha\vert V_B\rangle + e^{2\frac{\pi\:i}{3}}\beta\vert D_B\rangle + e^{4\frac{\pi\:i}{3}}\gamma\vert H_B\rangle) + \vert \Phi^1_{0AA}\rangle (\alpha\vert V_B\rangle + e^{2\pi\:i}\beta\vert D_B\rangle + e^{2\pi\:i}\gamma\vert H_B\rangle)\nonumber\\ && +\vert \Phi^2_{0AA}\rangle (\alpha\vert V_B\rangle + e^{4\frac{\pi\:i}{3}}\beta\vert D_B\rangle + e^{8\frac{\pi\:i}{3}}\gamma\vert H_B\rangle) + \vert \Phi^0_{1AA}\rangle (e^{2\frac{\pi\:i}{3}}\alpha\vert D_B\rangle + e^{4\frac{\pi\:i}{3}}\beta\vert H_B\rangle + \gamma\vert V_B\rangle) \nonumber\\ && + \vert \Phi^1_{1AA}\rangle (e^{2\frac{\pi\:i}{3}}\alpha\vert D_B\rangle + e^{8\frac{\pi\:i}{3}}\beta\vert H_B\rangle + e^{2\frac{\pi\:i}{3}}\gamma\vert V_B\rangle)+ \vert \Phi^2_{1AA}\rangle (e^{2\frac{\pi\:i}{3}}\alpha\vert D_B\rangle + e^{2\pi\:i}\beta\vert H_B\rangle + e^{4\frac{\pi\:i}{3}}\gamma\vert V_B\rangle) \nonumber\\ && + \vert \Phi^0_{2AA}\rangle (e^{4\frac{\pi\:i}{3}}\alpha\vert H_B\rangle + \beta\vert V_B\rangle + e^{2\frac{\pi\:i}{3}}\gamma\vert D_B\rangle) + \vert \Phi^1_{2AA}\rangle (e^{4\frac{\pi\:i}{3}}\alpha\vert H_B\rangle + e^{4\frac{\pi\:i}{3}}\beta\vert V_B\rangle + e^{4\frac{\pi\:i}{3}}\gamma\vert D_B\rangle) \nonumber\\ && + \vert \Phi^2_{2AA}\rangle (e^{4\frac{\pi\:i}{3}}\alpha\vert H_B\rangle + e^{2\frac{\pi\:i}{3}}\beta\vert V_B\rangle + e^{2\pi\:i}\gamma\vert D_B\rangle)\Big),
\end{eqnarray}
and
\begin{eqnarray}
\label{clubbedstate6}
\vert \Phi^{2}_{1AAB}\rangle^{cl} &=& \frac{1}{3}\Big(\vert \Phi^0_{0AA}\rangle (\alpha\vert V_B\rangle  + e^{4\frac{\pi\:i}{3}}\beta\vert D_B\rangle  + e^{2\frac{\pi\:i}{3}}\gamma\vert H_B\rangle ) + \vert \Phi^1_{0AA}\rangle (\alpha\vert V_B\rangle  + e^{8\frac{\pi\:i}{3}}\beta\vert D_B\rangle  + e^{4\frac{\pi\:i}{3}}\gamma\vert H_B\rangle )\nonumber\\ && +\vert \Phi^2_{0AA}\rangle (\alpha\vert V_B\rangle  + e^{2\pi\:i}\beta\vert D_B\rangle  + e^{2\pi\:i}\gamma\vert H_B\rangle ) + \vert \Phi^0_{1AA}\rangle (e^{4\frac{\pi\:i}{3}}\alpha\vert D_B\rangle  + e^{2\frac{\pi\:i}{3}}\beta\vert H_B\rangle  + \gamma\vert V_B\rangle ) \nonumber\\ && + \vert \Phi^1_{1AA}\rangle (e^{4\frac{\pi\:i}{3}}\alpha\vert D_B\rangle  + e^{2\pi\:i}\beta\vert H_B\rangle  + e^{2\frac{\pi\:i}{3}}\gamma\vert V_B\rangle )+ \vert \Phi^2_{1AA}\rangle (e^{4\frac{\pi\:i}{3}}\alpha\vert D_B\rangle  + e^{4\frac{\pi\:i}{3}}\beta\vert H_B\rangle  + e^{4\frac{\pi\:i}{3}}\gamma\vert V_B\rangle ) \nonumber\\ && + \vert \Phi^0_{2AA}\rangle (e^{2\frac{\pi\:i}{3}}\alpha\vert H_B\rangle + \beta\vert V_B\rangle + e^{4\frac{\pi\:i}{3}}\gamma\vert D_B\rangle ) + \vert \Phi^1_{2AA}\rangle (e^{2\frac{\pi\:i}{3}}\alpha\vert H_B\rangle  + e^{4\frac{\pi\:i}{3}}\beta\vert V_B\rangle  + e^{2\pi\:i}\gamma\vert D_B\rangle ) \nonumber\\ && + \vert \Phi^2_{2AA}\rangle (e^{2\frac{\pi\:i}{3}}\alpha\vert H_B\rangle  + e^{2\frac{\pi\:i}{3}}\beta\vert V_B\rangle  + e^{4\frac{\pi\:i}{3}}\gamma\vert D_B\rangle )\Big).
\end{eqnarray}
\noindent For each of the above cases shown as Eqs.(\ref{clubbedstate4})-(\ref{clubbedstate6}), two qutrits are possessed by Alice and a single qutrit is now being retained by Bob. Alice makes von-Neumann measurements on her qutrits and communicates her results to Bob. Consequently, Bob applies a set of unitary operators to retrieve the original single qutrit given in Eq.(\ref{singlequtrit}). Respectively, with respect to the Eqs.(\ref{clubbedstate4})-(\ref{clubbedstate6}), the unitary operators that Bob uses are $\Big\lbrace( V_{0}^{\prime})_{i}\Big\rbrace$, $\Big\lbrace( V_{0}^{\prime\prime})_{i}\Big\rbrace$, and $\Big\lbrace( V_{0}^{\prime\prime\prime})_{i}\Big\rbrace$, where $i = 0,1,\cdots, 8$. We get the following.

\subsubsection*{When $\vert \Phi^{0}_{1AB}\rangle$ is the qutrit channel:}

\begin{table}[h]
\caption{When the quantum channel is the state $\vert \Phi^{0}_{1AB}\rangle$}\label{tab3}%
\begin{tabular}{@{}ll@{}}
\toprule
\textbf{Measurement outcomes by Alice} & \textit{Unitary Operator that $B$ applies}\\
\midrule
$\vert \Phi^0_{0AA}\rangle$ & $(V_{0}^{\prime})_{0}\:(\alpha\vert V_B\rangle + \beta\vert D_B\rangle + \gamma\vert H_B\rangle)$\\

$\vert \Phi^1_{0AA}\rangle$ & $(V_{0}^{\prime})_{1}\:(\alpha\vert V_B\rangle + e^{4\frac{\pi\:i}{3}}\beta\vert D_B\rangle + e^{2\frac{\pi\:i}{3}}\gamma\vert H_B\rangle)$\\

$\vert \Phi^2_{0AA}\rangle$ & $(V_{0}^{\prime})_{2}\:(\alpha\vert V_B\rangle + e^{2\frac{\pi\:i}{3}}\beta\vert D_B\rangle + e^{4\frac{\pi\:i}{3}}\gamma\vert H_B\rangle)$\\

$\vert \Phi^0_{1AA}\rangle$ & $(V_{0}^{\prime})_{3}\:(\alpha\vert D_B\rangle + \beta\vert H_B\rangle + \gamma\vert V_B\rangle)$\\

$\vert \Phi^1_{1AA}\rangle$ & $(V_{0}^{\prime})_{4}\:(\alpha\vert D_B\rangle + e^{4\frac{\pi\:i}{3}}\beta\vert H_B\rangle + e^{2\frac{\pi\:i}{3}}\gamma\vert V_B\rangle)$\\

$\vert \Phi^2_{1AA}\rangle$ & $(V_{0}^{\prime})_{5}\:(\alpha\vert D_B\rangle + e^{2\frac{\pi\:i}{3}}\beta\vert H_B\rangle + e^{4\frac{\pi\:i}{3}}\gamma\vert V_B\rangle)$\\

$\vert \Phi^0_{2AA}\rangle$ & $(V_{0}^{\prime})_{6}\:(\alpha\vert H_B\rangle + \beta\vert V_B\rangle + \gamma\vert D_B\rangle)$\\

$\vert \Phi^1_{2AA}\rangle$ & $(V_{0}^{\prime})_{7}\:(\alpha\vert H_B\rangle + e^{4\frac{\pi\:i}{3}}\beta\vert V_B\rangle + e^{2\frac{\pi\:i}{3}}\gamma\vert D_B\rangle)$\\

$\vert \Phi^1_{2AA}\rangle$ & $(V_{0}^{\prime})_{8}\:(\alpha\vert H_B\rangle + e^{2\frac{\pi\:i}{3}}\beta\vert V_B\rangle + e^{4\frac{\pi\:i}{3}}\gamma\vert D_B\rangle)$\\
\hline
\botrule
\end{tabular}
\end{table}

\noindent Depending upon the outcomes obtained by Alice, when communicated to him, Bob applies appropriate unitary operators as shown in the above table $IV$. Here, $(V _{0}^{\prime})_{6}$ is the identity operator, which means that if Alice gets $\vert \Phi^0_{2}\rangle_{AA}$, then Bob is automatically left with the original single qutrit. He does nothing or we can say Bob applies identity operator $\mathcal{I}$ which is denoted here by  $(V_{0}^{\prime})_{6}$. Hence we get
\begin{eqnarray}
\label{unitary1}
(V_{0}^{\prime})_{0} &=& \vert H_B\rangle\langle V_B\vert  + \vert V_B\rangle\langle D_B\vert +  \vert D_B\rangle\langle H_B\vert,\nonumber\\
(V_{0}^{\prime})_{1} &=& \vert H_B\rangle\langle V_B\vert  + e^{2\frac{\pi\: i}{3}}\vert V_B\rangle\langle D_B\vert +  e^{4\frac{\pi\: i}{3}}\vert D_B\rangle\langle H_B\vert,\nonumber\\
(V_{0}^{\prime})_{2} &=& \vert H_B\rangle\langle V_B\vert  + e^{4\frac{\pi\: i}{3}}\vert V_B\rangle\langle D_B\vert +  e^{2\frac{\pi\: i}{3}}\vert D_B\rangle\langle H_B\vert,\nonumber\\
(V_{0}^{\prime})_{3} &=& \vert H_B\rangle\langle D_B\vert  + \vert V_B\rangle\langle H_B\vert +  \vert D_B\rangle\langle V_B\vert,\nonumber\\
(V_{0}^{\prime})_{4} &=& \vert H_B\rangle\langle D_B\vert  + e^{2\frac{\pi\: i}{3}}\vert V_B\rangle\langle H_B\vert +  e^{4\frac{\pi\: i}{3}}\vert D_B\rangle\langle V_B\vert,\nonumber\\
(V_{0}^{\prime})_{5} &=& \vert H_B\rangle\langle D_B\vert  + e^{2\frac{\pi\: i}{3}}\vert V_B\rangle\langle H_B\vert +  e^{4\frac{\pi\: i}{3}}\vert D_B\rangle\langle V_B\vert,\nonumber\\
(V_{0}^{\prime})_{6} &=& \mathcal{I}_B,\nonumber\\
(V_{0}^{\prime})_{7} &=& \vert H_B\rangle\langle H_B\vert + e^{2\frac{\pi\: i}{3}}\vert V_B\rangle\langle V_B\vert + e^{4\frac{\pi\: i}{3}}\vert D_B\rangle\langle D_B\vert, \nonumber\\
(V_{0}^{\prime})_{8} &=& \vert H_B\rangle\langle H_B\vert + e^{4\frac{\pi\: i}{3}}\vert V_B\rangle\langle V_B\vert + e^{2\frac{\pi\: i}{3}}\vert D_B\rangle\langle D_B\vert,.\nonumber\\
\end{eqnarray}
\subsubsection*{When $\vert \Phi^{1}_{1AB}\rangle$ is the qutrit channel:}

\begin{table}[h]
\caption{When the quantum channel is the state $\vert \Phi^{1}_{1AB}\rangle$}\label{tab3}%
\begin{tabular}{@{}ll@{}}
\toprule
\textbf{Measurement outcomes by Alice} & \textit{Unitary Operator that $B$ applies}\\
\midrule
$\vert \Phi^0_{0AA}\rangle$ & $(V_{0}^{\prime\prime})_{0}\:(\alpha\vert V_B\rangle + e^{2\frac{\pi\:i}{3}}\beta\vert D_B\rangle + e^{4\frac{\pi\:i}{3}}\gamma\vert H_B\rangle)$\\
$\vert \Phi^1_{0AA}\rangle$ & $(V_{0}^{\prime\prime})_{1}\:(\alpha\vert V_B\rangle + \beta\vert D_B\rangle  + \gamma\vert H_B\rangle )$\\
$\vert \Phi^2_{0AA}\rangle$ & $(V_{0}^{\prime\prime})_{2}\:(\alpha\vert V_B\rangle  + e^{4\frac{\pi\:i}{3}}\beta\vert D_B\rangle  + e^{8\frac{\pi\:i}{3}}\gamma\vert H_B\rangle )$\\
$\vert \Phi^0_{1AA}\rangle$ & $(V_{0}^{\prime\prime})_{3}\:(e^{2\frac{\pi\:i}{3}}\alpha\vert D_B\rangle  + e^{4\frac{\pi\:i}{3}}\beta\vert H_B\rangle  + \gamma\vert V_B\rangle )$\\
$\vert \Phi^1_{1AA}\rangle$ & $(V_{0}^{\prime\prime})_{4}\:(e^{2\frac{\pi\:i}{3}}\alpha\vert D_B\rangle + e^{8\frac{\pi\:i}{3}}\beta\vert H_B\rangle  + e^{2\frac{\pi\:i}{3}}\gamma\vert V_B\rangle )$\\
$\vert \Phi^2_{1AA}\rangle$ & $(V_{0}^{\prime\prime})_{5}\:(e^{2\frac{\pi\:i}{3}}\alpha\vert D_B\rangle  + \beta\vert H_B\rangle  + e^{4\frac{\pi\:i}{3}}\gamma\vert V_B\rangle )$\\
$\vert \Phi^0_{2AA}\rangle$ & $(V_{0}^{\prime\prime})_{6}\:(\alpha\vert H_B\rangle  + \beta\vert V_B\rangle  + e^{2\frac{\pi\:i}{3}}\gamma\vert D_B\rangle )$\\
$\vert \Phi^1_{2AA}\rangle$ & $(V_{0}^{\prime\prime})_{7}\:(\alpha\vert H_B\rangle  + e^{4\frac{\pi\:i}{3}}\beta\vert V_B\rangle  + e^{4\frac{\pi\:i}{3}}\gamma\vert D_B\rangle )$\\
$\vert \Phi^1_{2AA}\rangle$ & $(V_{0}^{\prime\prime})_{8}\:(e^{4\frac{\pi\:i}{3}}\alpha\vert H_B\rangle  + e^{2\frac{\pi\:i}{3}}\beta\vert V_B\rangle  + \gamma\vert D_B\rangle )$\\
\hline
\botrule
\end{tabular}
\end{table}
 
\noindent Bob applies appropriate unitary operators as shown in the above table $V$. Thus we get
\begin{eqnarray}
\label{unitary1}
(V_{0}^{\prime\prime})_{0} &=& \vert H_B\rangle\langle V_B\vert  + e^{4\frac{\pi\: i}{3}}\vert V_B\rangle\langle D_B\vert +  e^{2\frac{\pi\: i}{3}}\vert D_B\rangle\langle H_B\vert,\nonumber\\
(V_{0}^{\prime\prime})_{1} &=& \vert H_B\rangle\langle V_B\vert  + \vert V_B\rangle\langle D_B\vert +  \vert D_B\rangle\langle H_B\vert,\nonumber\\
(V_{0}^{\prime\prime})_{2} &=& \vert H_B\rangle\langle V_B\vert  + e^{2\frac{\pi\: i}{3}}\vert V_B\rangle\langle D_B\vert +  e^{-2\frac{\pi\: i}{3}}\vert D_B\rangle\langle H_B\vert,\nonumber\\
(V_{0}^{\prime\prime})_{3} &=& e^{4\frac{\pi\: i}{3}}\vert H_B\rangle\langle D_B\vert  + e^{2\frac{\pi\: i}{3}}\vert V_B\rangle\langle H_B\vert +  \vert D_B\rangle\langle V_B\vert,\nonumber\\
(V_{0}^{\prime\prime})_{4} &=& e^{4\frac{\pi\: i}{3}}\vert H_B\rangle\langle D_B\vert  + e^{-2\frac{\pi\: i}{3}}\vert V_B\rangle\langle H_B\vert +  e^{4\frac{\pi\: i}{3}}\vert D_B\rangle\langle V_B\vert,\nonumber\\
(V_{0}^{\prime\prime})_{5} &=& e^{4\frac{\pi\: i}{3}}\vert H_B\rangle\langle D_B\vert  + \vert V_B\rangle\langle H_B\vert +  e^{2\frac{\pi\: i}{3}}\vert D_B\rangle\langle V_B\vert,\nonumber\\
(V_{0}^{\prime\prime})_{6} &=& e^{2\frac{\pi\: i}{3}}\vert H_B\rangle\langle H_B\vert  + \vert V_B\rangle\langle V_B\vert +  e^{4\frac{\pi\: i}{3}}\vert D_B\rangle\langle D_B\vert,\nonumber\\
(V_{0}^{\prime\prime})_{7} &=& e^{2\frac{\pi\: i}{3}}\vert H_B\rangle\langle H_B\vert + e^{2\frac{\pi\: i}{3}}\vert V_B\rangle\langle V_B\vert + e^{2\frac{\pi\: i}{3}}\vert D_B\rangle\langle D_B\vert, \nonumber\\
(V_{0}^{\prime\prime})_{8} &=& e^{2\frac{\pi\: i}{3}}\vert H_B\rangle\langle H_B\vert + e^{4\frac{\pi\: i}{3}}\vert V_B\rangle\langle V_B\vert + \vert D_B\rangle\langle D_B\vert.
\end{eqnarray}
\subsubsection*{When $\vert \Phi^{2}_{1AB}\rangle$ is the qutrit channel:}

\begin{table}[h]
\caption{When the quantum channel is the state $\vert \Phi^{1}_{1AB}\rangle$}\label{tab3}%
\begin{tabular}{@{}ll@{}}
\toprule
\textbf{Measurement outcomes by Alice} & \textit{Unitary Operator that $B$ applies}\\
\midrule
$\vert \Phi^0_{0AA}\rangle$ & $(V_{0}^{\prime\prime\prime})_{0}\:(\alpha\vert V_B\rangle + e^{4\frac{\pi\:i}{3}}\beta\vert D_B\rangle + e^{2\frac{\pi\:i}{3}}\gamma\vert H_B\rangle)$\\
$\vert \Phi^1_{0AA}\rangle$ & $(V_{0}^{\prime\prime\prime})_{1}\:(\alpha\vert V_B\rangle  + e^{8\frac{\pi\:i}{3}}\beta\vert D_B\rangle  + e^{4\frac{\pi\:i}{3}}\gamma\vert H_B\rangle )$\\
$\vert \Phi^2_{0AA}\rangle$ & $(V_{0}^{\prime\prime\prime})_{2}\:(\alpha\vert V_B\rangle  + e^{2\pi\:i}\beta\vert D_B\rangle  + e^{2\pi\:i}\gamma\vert H_B\rangle )$\\
$\vert \Phi^0_{1AA}\rangle$ & $(V_{0}^{\prime\prime\prime})_{3}\:(e^{4\frac{\pi\:i}{3}}\alpha\vert D_B\rangle  + e^{2\frac{\pi\:i}{3}}\beta\vert H_B\rangle  + \gamma\vert V_B\rangle )$\\
$\vert \Phi^1_{1AA}\rangle$ & $(V_{0}^{\prime\prime\prime})_{4}\:(e^{4\frac{\pi\:i}{3}}\alpha\vert D_B\rangle  + e^{2\pi\:i}\beta\vert H_B\rangle  + e^{2\frac{\pi\:i}{3}}\gamma\vert V_B\rangle )$\\
$\vert \Phi^2_{1AA}\rangle$ & $(V_{0}^{\prime\prime\prime})_{5}\:(e^{4\frac{\pi\:i}{3}}\alpha\vert D_B\rangle  + e^{4\frac{\pi\:i}{3}}\beta\vert H_B\rangle  + e^{4\frac{\pi\:i}{3}}\gamma\vert V_B\rangle )$\\
$\vert \Phi^0_{2AA}\rangle$ & $(V_{0}^{\prime\prime\prime})_{6}\:(e^{2\frac{\pi\:i}{3}}\alpha\vert H_B\rangle  + \beta\vert V_B\rangle  + e^{4\frac{\pi\:i}{3}}\gamma\vert D_B\rangle )$\\
$\vert \Phi^1_{2AA}\rangle$ & $(V_{0}^{\prime\prime\prime})_{7}\:(e^{2\frac{\pi\:i}{3}}\alpha\vert H_B\rangle  + e^{4\frac{\pi\:i}{3}}\beta\vert V_B\rangle  + e^{2\pi\:i}\gamma\vert D_B\rangle )$\\
$\vert \Phi^1_{2AA}\rangle$ & $(V_{0}^{\prime\prime\prime})_{8}\:(e^{2\frac{\pi\:i}{3}}\alpha\vert H_B\rangle  + e^{2\frac{\pi\:i}{3}}\beta\vert V_B\rangle  + e^{4\frac{\pi\:i}{3}}\gamma\vert D_B\rangle )$\\
\hline
\botrule
\end{tabular}
\end{table}

\noindent Bob applies appropriate unitary operators as shown in the above table $VI$. Thus we get
\begin{eqnarray}
\label{unitary1}
(V_{0}^{\prime\prime\prime})_{0} &=& \vert H_B\rangle\langle V_B\vert  + e^{2\frac{\pi\: i}{3}}\vert V_B\rangle \langle D_B\vert +  e^{4\frac{\pi\: i}{3}}\vert D_B\rangle \langle H_B\vert,\nonumber\\
(V_{0}^{\prime\prime\prime})_{1} &=& \vert H_B\rangle \langle V_B\vert  + e^{-2\frac{\pi\: i}{3}}\vert V_B\rangle \langle D_B\vert +  e^{2\frac{\pi\: i}{3}}\vert D_B\rangle \langle H_B\vert,\nonumber\\
(V_{0}^{\prime\prime\prime})_{2} &=& \vert H_B\rangle \langle V_B\vert  + \vert V_B\rangle \langle D_B\vert +  \vert D_B\rangle \langle H_B\vert,\nonumber\\
(V_{0}^{\prime\prime\prime})_{3} &=& e^{2\frac{\pi\: i}{3}}\vert H_B\rangle \langle D_B\vert  + e^{4\frac{\pi\: i}{3}}\vert V_B\rangle \langle H_B\vert +  \vert D_B\rangle \langle V_B\vert,\nonumber\\
(V_{0}^{\prime\prime\prime})_{4} &=& e^{2\frac{\pi\: i}{3}}\vert H_B\rangle \langle D_B\vert  + \vert V_B\rangle \langle H_B\vert +  e^{4\frac{\pi\: i}{3}}\vert D_B\rangle \langle V_B\vert,\nonumber\\
(V_{0}^{\prime\prime\prime})_{5} &=& e^{2\frac{\pi\: i}{3}}\vert H_B\rangle \langle D_B\vert  + e^{2\frac{\pi\: i}{3}}\vert V_B\rangle \langle H_B\vert +  e^{2\frac{\pi\: i}{3}}\vert D_B\rangle \langle V_B\vert,\nonumber\\
(V_{0}^{\prime\prime\prime})_{6} &=& e^{4\frac{\pi\: i}{3}}\vert H_B\rangle \langle H_B\vert  + \vert V_B\rangle \langle V_B\vert +  e^{2\frac{\pi\: i}{3}}\vert D_B\rangle \langle D_B\vert,\nonumber\\
(V_{0}^{\prime\prime\prime})_{7} &=& e^{4\frac{\pi\: i}{3}}\vert H_B\rangle \langle H_B\vert + e^{2\frac{\pi\: i}{3}}\vert V_B\rangle \langle V_B\vert + \vert D_B\rangle \langle D_B\vert, \nonumber\\
(V_{0}^{\prime\prime\prime})_{8} &=& e^{4\frac{\pi\: i}{3}}\vert H_B\rangle \langle H_B\vert + e^{4\frac{\pi\: i}{3}}\vert V_B\rangle \langle V_B\vert + e^{2\frac{\pi\: i}{3}}\vert D_B\rangle \langle D_B\vert.
\end{eqnarray}
\subsection*{The states $\vert \Phi^{0}_{2AB}\rangle,\:\vert \Phi^{1}_{2AB}\rangle,\:\vert \Phi^{2}_{2AB}\rangle$ as teleportation channels:}
\noindent Again if we consider the states defined in Eqs.(\ref{eq:sys:g})-(\ref{eq:sys:i})as quantum channels for teleportation, then using Eqs.(\ref{rewritebasis}) the following are obtained, when Alice clubs her single qutrit to the channels.  Respectively, with respect to the Eqs.(\ref{clubbedstate7})-(\ref{clubbedstate9}), the unitary operators that Bob uses are $\Big\lbrace( W_{0}^{\prime})_{i}\Big\rbrace$, $\Big\lbrace( W_{0}^{\prime\prime})_{i}\Big\rbrace$, and $\Big\lbrace( W_{0}^{\prime\prime\prime})_{i}\Big\rbrace$, where $i = 0,1,\cdots, 8$. We get the following.
\begin{eqnarray}
\label{clubbedstate7}
\vert \Phi^{0}_{2AAB}\rangle &=& \frac{1}{3}\Big(\vert \Phi^0_{0AA}\rangle (\alpha\vert D_B\rangle  + \beta\vert H_B\rangle  + \gamma\vert V_B\rangle) + \vert \Phi^1_{0AA}\rangle (\alpha\vert D_B\rangle  + e^{4\frac{\pi\:i}{3}}\beta\vert H_B\rangle  + e^{2\frac{\pi\:i}{3}}\gamma\vert V_B\rangle )\nonumber\\ && +\vert \Phi^2_{0AA}\rangle (\alpha\vert D_B\rangle + e^{2\frac{\pi\:i}{3}}\beta\vert H_B\rangle  + e^{4\frac{\pi\:i}{3}}\gamma\vert V_B\rangle ) + \vert \Phi^0_{1AA}\rangle (\alpha\vert H_B\rangle  + \beta\vert V_B\rangle  + \gamma\vert D_B\rangle ) \nonumber\\ && + \vert \Phi^1_{1AA}\rangle (\alpha\vert H_B\rangle  + e^{4\frac{\pi\:i}{3}}\beta\vert V_B\rangle  + e^{2\frac{\pi\:i}{3}}\gamma\vert D_B\rangle )+ \vert \Phi^2_{1AA}\rangle (\alpha\vert H_B\rangle  + e^{2\frac{\pi\:i}{3}}\beta\vert V_B\rangle  + e^{4\frac{\pi\:i}{3}}\gamma\vert D_B\rangle ) \nonumber\\ && + \vert \Phi^0_{2AA}\rangle (\alpha\vert V_B\rangle  + \beta\vert D_B\rangle  + \gamma\vert H_B\rangle ) + \vert \Phi^1_{2AA}\rangle (\alpha\vert V_B\rangle  + e^{4\frac{\pi\:i}{3}}\beta\vert D_B\rangle  + e^{2\frac{\pi\:i}{3}}\gamma\vert H_B\rangle ) \nonumber\\ && + \vert \Phi^2_{2AA}\rangle (\alpha\vert V_B\rangle  + e^{2\frac{\pi\:i}{3}}\beta\vert D_B\rangle  + e^{4\frac{\pi\:i}{3}}\gamma\vert H_B\rangle )\Big),
\end{eqnarray}
\begin{eqnarray}
\label{clubbedstate8}
\vert \Phi^{1}_{2AAB}\rangle &=& \frac{1}{3}\Big(\vert \Phi^0_{0AA}\rangle (\alpha\vert D_B\rangle + e^{2\frac{\pi\:i}{3}}\beta\vert H_B\rangle  + e^{4\frac{\pi\:i}{3}}\gamma\vert V_B\rangle ) + \vert \Phi^1_{0AA}\rangle (\alpha\vert D_B\rangle  + e^{2\pi\:i}\beta\vert H_B\rangle  + e^{2\pi\:i}\gamma\vert V_B\rangle )\nonumber\\ && +\vert \Phi^2_{0AA}\rangle (\alpha\vert D_B\rangle + e^{4\frac{\pi\:i}{3}}\beta\vert H_B\rangle  + e^{8\frac{\pi\:i}{3}}\gamma\vert V_B\rangle ) + \vert \Phi^0_{1AA}\rangle (e^{2\frac{\pi\:i}{3}}\alpha\vert H_B\rangle  + e^{4\frac{\pi\:i}{3}}\beta\vert V_B\rangle  + \gamma\vert D_B\rangle ) \nonumber\\ && + \vert \Phi^1_{1AA}\rangle (e^{2\frac{\pi\:i}{3}}\alpha\vert H_B\rangle  + e^{8\frac{\pi\:i}{3}}\beta\vert V_B\rangle  + e^{2\frac{\pi\:i}{3}}\gamma\vert D_B\rangle )+ \vert \Phi^2_{1}\rangle (e^{2\frac{\pi\:i}{3}}\alpha\vert H_B\rangle  + e^{2\pi\:i}\beta\vert V_B\rangle  + e^{4\frac{\pi\:i}{3}}\gamma\vert D_B\rangle ) \nonumber\\ && + \vert \Phi^0_{2AA}\rangle (e^{4\frac{\pi\:i}{3}}\alpha\vert V_B\rangle  + \beta\vert D_B\rangle  + e^{2\frac{\pi\:i}{3}}\gamma\vert H_B\rangle ) + \vert \Phi^1_{2AA}\rangle (e^{4\frac{\pi\:i}{3}}\alpha\vert V_B\rangle  + e^{4\frac{\pi\:i}{3}}\beta\vert D_B\rangle  + e^{4\frac{\pi\:i}{3}}\gamma\vert H_B\rangle ) \nonumber\\ && + \vert \Phi^2_{2AA}\rangle (e^{4\frac{\pi\:i}{3}}\alpha\vert V_B\rangle  + e^{2\frac{\pi\:i}{3}}\beta\vert D_B\rangle  + e^{2\pi\:i}\gamma\vert H_B\rangle )\Big),
\end{eqnarray}
and
\begin{eqnarray}
\label{clubbedstate9}
\vert \Phi^{2}_{2AAB}\rangle &=& \frac{1}{3}\Big(\vert \Phi^0_{0AA}\rangle (\alpha\vert D_B\rangle  + e^{4\frac{\pi\:i}{3}}\beta\vert H_B\rangle  + e^{2\frac{\pi\:i}{3}}\gamma\vert V_B\rangle ) + \vert \Phi^1_{0AA}\rangle (\alpha\vert D_B\rangle  + e^{8\frac{\pi\:i}{3}}\beta\vert H_B\rangle  + e^{4\frac{\pi\:i}{3}}\gamma\vert V_B\rangle )\nonumber\\ && +\vert \Phi^2_{0AA}\rangle (\alpha\vert D_B\rangle  + e^{2\pi\:i}\beta\vert H_B\rangle  + e^{2\pi\:i}\gamma\vert V_B\rangle ) + \vert \Phi^0_{1AA}\rangle (e^{4\frac{\pi\:i}{3}}\alpha\vert H_B\rangle  + e^{2\frac{\pi\:i}{3}}\beta\vert V_B\rangle  + \gamma\vert D_B\rangle ) \nonumber\\ && + \vert \Phi^1_{1AA}\rangle (e^{4\frac{\pi\:i}{3}}\alpha\vert H_B\rangle  + e^{2\pi\:i}\beta\vert V_B\rangle  + e^{2\frac{\pi\:i}{3}}\gamma\vert D_B\rangle )+ \vert \Phi^2_{1}\rangle (e^{4\frac{\pi\:i}{3}}\alpha\vert H_B\rangle  + e^{4\frac{\pi\:i}{3}}\beta\vert V_B\rangle  + e^{4\frac{\pi\:i}{3}}\gamma\vert D_B\rangle ) \nonumber\\ && + \vert \Phi^0_{2AA}\rangle (e^{2\frac{\pi\:i}{3}}\alpha\vert V_B\rangle  + \beta\vert D_B\rangle  + e^{4\frac{\pi\:i}{3}}\gamma\vert H_B\rangle ) + \vert \Phi^1_{2AA}\rangle (e^{2\frac{\pi\:i}{3}}\alpha\vert V_B\rangle  + e^{4\frac{\pi\:i}{3}}\beta\vert D_B\rangle  + e^{2\pi\:i}\gamma\vert H_B\rangle ) \nonumber\\ && + \vert \Phi^2_{2AA}\rangle (e^{2\frac{\pi\:i}{3}}\alpha\vert V_B\rangle + e^{2\frac{\pi\:i}{3}}\beta\vert D_B\rangle  + e^{8\frac{\pi\:i}{3}}\gamma\vert H_B\rangle )\Big).
\end{eqnarray}
\subsubsection*{When $\vert \Phi^{0}_{2AB}\rangle$ is the qutrit channel:}

\begin{table}[h]
\caption{When the quantum channel is the state $\vert \Phi^{0}_{2AB}\rangle$}\label{tab3}%
\begin{tabular}{@{}ll@{}}
\toprule
\textbf{Measurement outcomes by Alice} & \textit{Unitary Operator that $B$ applies}\\
\midrule
$\vert \Phi^0_{0AA}\rangle$ & $(W_{0}^{\prime})_{0}\:(\alpha\vert D_B\rangle + \beta\vert H_B\rangle  + \gamma\vert V_B\rangle )$\\
$\vert \Phi^1_{0AA}\rangle$ & $(W_{0}^{\prime})_{1}\:(\alpha\vert D_B\rangle  + e^{4\frac{\pi\:i}{3}}\beta\vert H_B\rangle  + e^{2\frac{\pi\:i}{3}}\gamma\vert V_B\rangle )$\\
$\vert \Phi^2_{0AA}\rangle$ & $(W_{0}^{\prime})_{2}\:(\alpha\vert D_B\rangle + e^{2\frac{\pi\:i}{3}}\beta\vert H_B\rangle  + e^{4\frac{\pi\:i}{3}}\gamma\vert V_B\rangle )$\\
$\vert \Phi^0_{1AA}\rangle$ & $(W_{0}^{\prime})_{3}\:(\alpha\vert H_B\rangle  + \beta\vert V_B\rangle  + \gamma\vert D_B\rangle )$\\
$\vert \Phi^1_{1AA}\rangle$ & $(W_{0}^{\prime})_{4}\:(\alpha\vert H_B\rangle  + e^{4\frac{\pi\:i}{3}}\beta\vert V_B\rangle  + e^{2\frac{\pi\:i}{3}}\gamma\vert D_B\rangle )$\\
$\vert \Phi^2_{1AA}\rangle$ & $(W_{0}^{\prime})_{5}\:(\alpha\vert H_B\rangle  + e^{2\frac{\pi\:i}{3}}\beta\vert V_B\rangle  + e^{4\frac{\pi\:i}{3}}\gamma\vert D_B\rangle )$\\
$\vert \Phi^0_{2AA}\rangle$ & $(W_{0}^{\prime})_{6}\:(\alpha\vert V_B\rangle  + \beta\vert D_B\rangle  + \gamma\vert H_B\rangle )$\\
$\vert \Phi^1_{2AA}\rangle$ & $(W_{0}^{\prime})_{7}\:(\alpha\vert V_B\rangle  + e^{4\frac{\pi\:i}{3}}\beta\vert D_B\rangle  + e^{2\frac{\pi\:i}{3}}\gamma\vert H_B\rangle )$\\
$\vert \Phi^1_{2AA}\rangle$ & $(W_{0}^{\prime})_{8}\:(\alpha\vert V_B\rangle  + e^{2\frac{\pi\:i}{3}}\beta\vert D_B\rangle  + e^{4\frac{\pi\:i}{3}}\gamma\vert H_B\rangle )$\\
\hline
\botrule
\end{tabular}
\end{table}

\noindent Bob applies appropriate unitary operators as shown in the above table $VII$. Thus we get
\begin{eqnarray}
\label{unitary1}
(W_{0}^{\prime})_{0} &=& \vert H_B\rangle\langle D_B\vert  + \vert V_B\rangle\langle H_B\vert +  \vert D_B\rangle\langle V_B\vert,\nonumber\\
(W_{0}^{\prime})_{1} &=& \vert H_B\rangle\langle D_B\vert  + e^{4\frac{\pi\: i}{3}}\vert V_B\rangle\langle H_B\vert +  e^{2\frac{\pi\: i}{3}}\vert D_B\rangle\langle V_B\vert,\nonumber\\
(W_{0}^{\prime})_{2} &=& \vert H_B\rangle\langle D_B\vert  + e^{2\frac{\pi\: i}{3}}\vert V_B\rangle\langle H_B\vert +  e^{4\frac{\pi\: i}{3}}\vert D_B\rangle\langle V_B\vert,\nonumber\\
(W_{0}^{\prime})_{3} &=& \mathcal{I}_B,\nonumber\\
(W_{0}^{\prime})_{4} &=& \vert H_B\rangle\langle H_B\vert  + e^{2\frac{\pi\: i}{3}}\vert V_B\rangle\langle V_B\vert +  e^{4\frac{\pi\: i}{3}}\vert D_B\rangle\langle D_B\vert,\nonumber\\
(W_{0}^{\prime})_{5} &=& \vert H_B\rangle\langle H_B\vert  + e^{4\frac{\pi\: i}{3}}\vert V_B\rangle\langle V_B\vert +  e^{2\frac{\pi\: i}{3}}\vert D_B\rangle\langle D_B\vert,\nonumber\\
(W_{0}^{\prime})_{6} &=& \vert H_B\rangle\langle V_B\vert  + \vert V_B\rangle\langle D_B\vert +  \vert D_B\rangle\langle H_B\vert,\nonumber\\
(W_{0}^{\prime})_{7} &=& \vert H_B\rangle\langle V_B\vert + e^{2\frac{\pi\: i}{3}}\vert V_B\rangle\langle D_B\vert + e^{4\frac{\pi\: i}{3}}\vert D_B\rangle\langle H_B\vert, \nonumber\\
(W_{0}^{\prime})_{8} &=& \vert H_B\rangle\langle V_B\vert + e^{4\frac{\pi\: i}{3}}\vert V_B\rangle\langle D_B\vert + e^{2\frac{\pi\: i}{3}}\vert D_B\rangle\langle H_B\vert.
\end{eqnarray}

\subsubsection*{When $\vert \Phi^{1}_{2AB}\rangle$ is the qutrit channel:}

\begin{table}[h]
\caption{When the quantum channel is the state $\vert \Phi^{0}_{2AB}\rangle$}\label{tab3}%
\begin{tabular}{@{}ll@{}}
\toprule
\textbf{Measurement outcomes by Alice} & \textit{Unitary Operator that $B$ applies}\\
\midrule
$\vert \Phi^0_{0AA}\rangle $ & $(W_{0}^{\prime\prime})_{0}\:(\alpha\vert D_B\rangle  + e^{2\frac{\pi\:i}{3}}\beta\vert H_B\rangle  + e^{4\frac{\pi\:i}{3}}\gamma\vert V_B\rangle_{B})$\\
$\vert \Phi^1_{0AA}\rangle $ & $(W_{0}^{\prime\prime})_{1}\:(\alpha\vert D_B\rangle  + e^{2\pi\:i}\beta\vert H_B\rangle  + e^{2\pi\:i}\gamma\vert V_B\rangle )$\\
$\vert \Phi^2_{0AA}\rangle $ & $(W_{0}^{\prime\prime})_{2}\:(\alpha\vert D_B\rangle + e^{4\frac{\pi\:i}{3}}\beta\vert H_B\rangle  + e^{8\frac{\pi\:i}{3}}\gamma\vert V_B\rangle )$\\
$\vert \Phi^0_{1AA}\rangle $ & $(W_{0}^{\prime\prime})_{3}\:(e^{2\frac{\pi\:i}{3}}\alpha\vert H_B\rangle  + e^{4\frac{\pi\:i}{3}}\beta\vert V_B\rangle  + \gamma\vert D_B\rangle )$\\
$\vert \Phi^1_{1AA}\rangle $ & $(W_{0}^{\prime\prime})_{4}\:(e^{2\frac{\pi\:i}{3}}\alpha\vert H_B\rangle  + e^{8\frac{\pi\:i}{3}}\beta\vert V_B\rangle  + e^{2\frac{\pi\:i}{3}}\gamma\vert D_B\rangle )$\\
$\vert \Phi^2_{1AA}\rangle $ & $(W_{0}^{\prime\prime})_{5}\:(e^{2\frac{\pi\:i}{3}}\alpha\vert H_B\rangle + e^{2\pi\:i}\beta\vert V_B\rangle  + e^{4\frac{\pi\:i}{3}}\gamma\vert D_B\rangle )$\\
$\vert \Phi^0_{2AA}\rangle $ & $(W_{0}^{\prime\prime})_{6}\:(e^{4\frac{\pi\:i}{3}}\alpha\vert V_B\rangle  + \beta\vert D_B\rangle  + e^{2\frac{\pi\:i}{3}}\gamma\vert H_B\rangle )$\\
$\vert \Phi^1_{2AA}\rangle $ & $(W_{0}^{\prime\prime})_{7}\:(e^{4\frac{\pi\:i}{3}}\alpha\vert V_B\rangle  + e^{4\frac{\pi\:i}{3}}\beta\vert D_B\rangle  + e^{4\frac{\pi\:i}{3}}\gamma\vert H_B\rangle )$\\
$\vert \Phi^1_{2AA}\rangle $ & $(W_{0}^{\prime\prime})_{8}\:(e^{4\frac{\pi\:i}{3}}\alpha\vert V_B\rangle  + e^{2\frac{\pi\:i}{3}}\beta\vert D_B\rangle  + e^{2\pi\:i}\gamma\vert H_B\rangle )$\\
\hline
\botrule
\end{tabular}
\end{table}

\noindent Bob applies appropriate unitary operators as shown in the above table~$VIII$. Thus we get
\begin{eqnarray}
\label{unitary1}
(W_{0}^{\prime\prime})_{0} &=& \vert H_B\rangle \langle D_B\vert  + e^{4\frac{\pi\: i}{3}}\vert V_B\rangle \langle H_B\vert +  e^{2\frac{\pi\: i}{3}}\vert D_B\rangle \langle V_B\vert,\nonumber\\
(W_{0}^{\prime\prime})_{1} &=& \vert H_B\rangle \langle D_B\vert  + \vert V_B\rangle \langle H_B\vert +  \vert D_B\rangle \langle V_B\vert,\nonumber\\
(W_{0}^{\prime\prime})_{2} &=& \vert H_B\rangle \langle D_B\vert  + e^{2\frac{\pi\: i}{3}}\vert V_B\rangle \langle H_B\vert +  e^{-2\frac{\pi\: i}{3}}\vert D_B\rangle \langle V_B\vert,\nonumber\\
(W_{0}^{\prime\prime})_{3} &=& e^{4\frac{\pi\: i}{3}}\vert H_B\rangle \langle H_B\vert  + e^{2\frac{\pi\: i}{3}}\vert V_B\rangle \langle V_B\vert +  \vert D_B\rangle \langle D_B\vert ,\nonumber\\
(W_{0}^{\prime\prime})_{4} &=& e^{4\frac{\pi\: i}{3}}\vert H_B\rangle \langle H_B\vert  + e^{-2\frac{\pi\: i}{3}}\vert V_B\rangle \langle V_B\vert +  e^{4\frac{\pi\: i}{3}}\vert D_B\rangle \langle D_B\vert,\nonumber\\
(W_{0}^{\prime\prime})_{5} &=& e^{4\frac{\pi\: i}{3}}\vert H_B\rangle \langle H_B\vert  +\vert V_B\rangle \langle V_B\vert +  e^{2\frac{\pi\: i}{3}}\vert D_B\rangle \langle D_B\vert,\nonumber\\
(W_{0}^{\prime\prime})_{6} &=& e^{2\frac{\pi\: i}{3}}\vert H_B\rangle \langle V_B\vert  + \vert V_B\rangle \langle D_B\vert +  e^{4\frac{\pi\: i}{3}}\vert D_B\rangle \langle H_B\vert,\nonumber\\
(W_{0}^{\prime\prime})_{7} &=& e^{2\frac{\pi\: i}{3}}\vert H_B\rangle \langle V_B\vert + e^{2\frac{\pi\: i}{3}}\vert V_B\rangle \langle D_B\vert + e^{2\frac{\pi\: i}{3}}\vert D_B\rangle \langle H_B\vert, \nonumber\\
(W_{0}^{\prime\prime})_{8} &=& e^{2\frac{\pi\: i}{3}}\vert H_B\rangle \langle V_B\vert + e^{4\frac{\pi\: i}{3}}\vert V_B\rangle \langle D_B\vert + \vert D_B\rangle \langle H_B\vert.
\end{eqnarray}
\subsubsection*{When $\vert \Phi^{2}_{2AB}\rangle$ is the qutrit channel:}

\begin{table}[h]
\caption{When the quantum channel is the state $\vert \Phi^{2}_{2AB}\rangle$}\label{tab3}%
\begin{tabular}{@{}ll@{}}
\toprule
\textbf{Measurement outcomes by Alice} & \textit{Unitary Operator that $B$ applies}\\
\midrule
$\vert \Phi^0_{0AA}\rangle$ & $(W_{0}^{\prime\prime\prime})_{0}\:(\alpha\vert D_B\rangle + e^{4\frac{\pi\:i}{3}}\beta\vert H_B\rangle + e^{2\frac{\pi\:i}{3}}\gamma\vert V_B\rangle)$\\
$\vert \Phi^1_{0AA}\rangle$ & $(W_{0}^{\prime\prime\prime})_{1}\:(\alpha\vert D_B\rangle + e^{8\frac{\pi\:i}{3}}\beta\vert H_B\rangle + e^{4\frac{\pi\:i}{3}}\gamma\vert V_B\rangle)$\\
$\vert \Phi^2_{0AA}\rangle$ & $(W_{0}^{\prime\prime\prime})_{2}\:(\alpha\vert D_B\rangle  + e^{2\pi\:i}\beta\vert H_B\rangle  + e^{2\pi\:i}\gamma\vert V_B\rangle )$\\
$\vert \Phi^0_{1AA}\rangle$ & $(W_{0}^{\prime\prime\prime})_{3}\:(e^{4\frac{\pi\:i}{3}}\alpha\vert H_B\rangle  + e^{2\frac{\pi\:i}{3}}\beta\vert V_B\rangle  + \gamma\vert D_B\rangle )$\\
$\vert \Phi^1_{1AA}\rangle $ & $(W_{0}^{\prime\prime\prime})_{4}\:(e^{4\frac{\pi\:i}{3}}\alpha\vert H_B\rangle  + e^{2\pi\:i}\beta\vert V_B\rangle  + e^{2\frac{\pi\:i}{3}}\gamma\vert D_B\rangle )$\\
$\vert \Phi^2_{1AA}\rangle $ & $(W_{0}^{\prime\prime\prime})_{5}\:(e^{4\frac{\pi\:i}{3}}\alpha\vert H_B\rangle  + e^{4\frac{\pi\:i}{3}}\beta\vert V_B\rangle  + e^{4\frac{\pi\:i}{3}}\gamma\vert D_B\rangle )$\\
$\vert \Phi^0_{2AA}\rangle $ & $(W_{0}^{\prime\prime\prime})_{6}\:(e^{2\frac{\pi\:i}{3}}\alpha\vert V_B\rangle  + \beta\vert D_B\rangle + e^{4\frac{\pi\:i}{3}}\gamma\vert H_B\rangle )$\\
$\vert \Phi^1_{2AA}\rangle$ & $(W_{0}^{\prime\prime\prime})_{7}\:(\alpha\vert V_B\rangle  + e^{4\frac{\pi\:i}{3}}\beta\vert D_B\rangle  + e^{2\frac{\pi\:i}{3}}\gamma\vert H_B\rangle )$\\
$\vert \Phi^1_{2AA}\rangle $ & $(W_{0}^{\prime\prime\prime})_{8}\:(e^{2\frac{\pi\:i}{3}}\alpha\vert V_B\rangle  + e^{2\frac{\pi\:i}{3}}\beta\vert D_B\rangle  + e^{8\frac{\pi\:i}{3}}\gamma\vert H_B\rangle )$\\
\hline
\botrule
\end{tabular}
\end{table}

\noindent Bob applies appropriate unitary operators as shown in the above table \ref{table8}. Thus we get
\begin{eqnarray}
\label{unitary1}
(W_{0}^{\prime\prime\prime})_{0} &=& \vert H_B\rangle\langle D_B\vert  + e^{2\frac{\pi\: i}{3}}\vert V_B\rangle_{B}\langle H_B\vert +  e^{4\frac{\pi\: i}{3}}\vert D_B\rangle_{B}\langle V_B\vert,\nonumber\\
(W_{0}^{\prime\prime\prime})_{1} &=& \vert H_B\rangle_{B}\langle D_B\vert  + e^{-2\frac{\pi\: i}{3}}\vert V_B\rangle_{B}\langle H_B\vert +  e^{2\frac{\pi\: i}{3}}\vert D_B\rangle_{B}\langle V_B\vert,\nonumber\\
(W_{0}^{\prime\prime\prime})_{2} &=& \vert H_B\rangle\langle D_B\vert  + \vert V_B\rangle\langle H_B\vert + \vert D_B\rangle\langle V_B\vert,\nonumber\\
(W_{0}^{\prime\prime\prime})_{3} &=& e^{2\frac{\pi\: i}{3}}\vert H_B\rangle \langle H_B\vert  + e^{4\frac{\pi\: i}{3}}\vert V_B\rangle \langle V_B\vert +  \vert D_B\rangle \langle D_B\vert ,\nonumber\\
(W_{0}^{\prime\prime\prime})_{4} &=& e^{2\frac{\pi\: i}{3}}\vert H_B\rangle \langle H_B\vert  + \vert V_B\rangle \langle V_B\vert +  e^{4\frac{\pi\: i}{3}}\vert D_B\rangle \langle D_B\vert,\nonumber\\
(W_{0}^{\prime\prime\prime})_{5} &=& e^{2\frac{\pi\: i}{3}}\vert H_B\rangle \langle H_B\vert  +e^{2\frac{\pi\: i}{3}}\vert V_B\rangle \langle V_B\vert +  e^{2\frac{\pi\: i}{3}}\vert D_B\rangle\langle D_B\vert,\nonumber\\
(W_{0}^{\prime\prime\prime})_{6} &=& e^{4\frac{\pi\: i}{3}}\vert H_B\rangle\langle V_B\vert  + \vert V_B\rangle\langle D_B\vert +  e^{2\frac{\pi\: i}{3}}\vert D_B\rangle\langle H_B\vert,\nonumber\\
(W_{0}^{\prime\prime\prime})_{7} &=& e^{4\frac{\pi\: i}{3}}\vert H_B\rangle\langle V_B\vert + e^{2\frac{\pi\: i}{3}}\vert V_B\rangle\langle D_B\vert + \vert D_B\rangle\langle H_B\vert, \nonumber\\
(W_{0}^{\prime\prime\prime})_{8} &=& e^{4\frac{\pi\: i}{3}}\vert H_B\rangle\langle V_B\vert + e^{4\frac{\pi\: i}{3}}\vert V_B\rangle\langle D_B\vert + e^{-2\frac{\pi\: i}{3}}\vert D_B\rangle\langle H_B\vert.
\end{eqnarray}
\section{Conclusion:}

\noindent This paper presents a comprehensive and novel protocol for the deterministic teleportation of a single qutrit utilizing two-qutrit entangled states constructed from symmetric and anti-symmetric basis states. By leveraging the unique properties of these specially constructed qutrit basis states, the protocol enables the faithful and efficient teleportation of a single qutrit state between two parties, traditionally denoted as Alice and Bob. Our scheme extends the foundational principles of quantum teleportation beyond qubit systems to higher-dimensional quantum systems, thereby harnessing the increased information capacity and enhanced security features that qutrits inherently offer. We have systematically demonstrated the use of nine distinct two-qutrit entangled states as quantum channels for teleportation. For each channel, detailed unitary operations required at the receiver's end have been explicitly formulated to recover the original unknown qutrit state after the sender performs the appropriate joint measurements and communicates the outcomes via classical channels. This rigorous approach confirms the deterministic nature of the teleportation process with perfect fidelity. By adopting the symmetric and anti-symmetric qutrit states proposed by Leslie et al., our work introduces new viable resource states for quantum communication tasks. The ability to teleport single qutrits deterministically using these structured entangled states substantially enriches the toolkit available for quantum information processing, especially in scenarios demanding higher-dimensional encoding. Such advancements are expected to enhance the development of quantum networks, quantum computing algorithms, and quantum cryptography schemes by exploiting the richer state space offered by qutrits. Furthermore, this study provides valuable insights into the behavior of symmetric and anti-symmetric quantum states and their practical utility in quantum information science. The framework laid out here can be extended to even larger multipartite systems, opening up avenues for future research into high-dimensional entanglement and its applications. In conclusion, our work not only advances the theoretical understanding of qutrit-based quantum teleportation but also has potential implications for the experimental realization of deterministic higher-dimensional quantum communication. The results presented here pave the way for more complex quantum information protocols utilizing qutrits and contribute to bridging the gap between theory and practical quantum technologies.

\noindent 
\vskip 0.5cm
{\bf Declaration of competing interest} The authors declare that they have no known competing financial interests or
personal relationships that could have appeared to influence the work reported in this paper.
\vskip 0.5cm
{\bf Data availability statement} All data that support the findings of this study are included within the article. No supplementary file has been added.
\section{Acknowledgement:} 
The authors acknowledge Prof. Tushar K. Dey and Prof. Surajit Sen (of Centre of Advanced Studies and Innovation Lab,
18/27, Tarapur, Silchar, INDIA) for the valuable comments.

\bibliography{Bibliography/refllm}
\end{document}